\documentclass[aps,prl,twocolumn,superscriptaddress,showpacs]{revtex4}
\usepackage[dvips]{graphicx}

\begin{document}

\title{Quantum Theory of Spontaneous Emission from Exciton-Electron-Phonon Complexes in Solid: Quantum Interference and Many-Body Effect
}

\author{Shi-Jie Xiong}
\email{sjxiong@nju.edu.cn}
\address{National Laboratory of Solid State Microstructures and
Department of Physics, Nanjing University, Nanjing 210093, China}
\address{Department of Physics and HKU-CAS Joint Laboratory on
New Materials, The University of Hong Kong, Pokfulam Road, Hong Kong, China}
\author{S. J. Xu}
\email{sjxu@hkucc.hku.hk}
\address{Department of Physics and HKU-CAS Joint Laboratory on
New Materials, The University of Hong Kong, Pokfulam Road, Hong Kong, China}

\begin{abstract}

A full quantum mechanical theory for the spontaneous emission from excitons simultaneously coupled to electronic excitations and anharmonic phonons in solid is developed. Origin of detailed structures, such as zero-phonon line splitting, Fano lineshape near one phonon sideband, strong second-order phonon Stokes line, asymmetric phonon anti-Stokes lines, and two-electron satellites as well as their phonon replicas recently revealed in the low-temperature photoluminescence of ZnO, has been identified by quantitative calculations from the theory.
\end{abstract}


\maketitle


Exciton, a bound electron-hole pair which does not carry electric current but performs various correlated motion of electron and hole, is of fundamental importance in optical properties of solids.\cite{aa1,bb1,cc1} Over 70 year development has made the exciton physics become a highly diversified field of science and is still being actively pursued in many directions since excitons have been found in all types of non-metallic solids as well as in certain rare earth metals. The interactions of excitons with each other and with other quasi-particles such as phonons drastically affect various aspects of the exciton related phenomena. Furthermore, the subject is intimately connected with many other fields of physics. For example, the neutral-donor-bound excitons in solids have been investigated as one of promising building blocks for future quantum computer.\cite{y}

It has been well established that the coupling of excitons with phonons leads to the phonon
sidebands in absorption or emission spectra which reflect frequencies of phonons and the
coupling strength \cite{aa2,aa3}. On the other hand, interactions of excitons
with electronic excitations can also be revealed by lineshapes
and radiation frequencies in the photoemission spectra \cite{b1,b2,b3}.
Theoretically, the influences of interactions of excitons with
phononic and electronic excitations are usually treated separately
by perturbation theories together with Green function technique
\cite{tt1,tt2}. In real solids, however, excitons are often simultaneously coupled to surrounding electronic and phononic excitations \cite{exp}. There are still difficulties to theoretically handle the combined effects of 
interactions with both phonons and electrons. Especially, the
quantum coherence recently discovered in spontaneous emissions eagerly demands a firm theoretical identification.

In this Letter we develop an approach to calculate the spontaneous emission spectrum of excitons interacting with electrons and anharmonic phonons on the basis of exact solutions of
excitonic many-body states. By this method the
energy distributions of different excitations, the effect of quantum
interference among components, and the many-body effect in exciton-electron-phonon coupling systems, are taken into account on an equal footing. It is found that detailed spectral features can be quantitatively explained by the theory.

We consider one quantum system containing subsystems of exciton, phonon and electron. In uncoupled case they are expressed respectively by $H_{ex} = \epsilon_0
a^\dag a$, $H_{ph}=\hbar \omega_{0}
b^{\dag}_{0} b_{0}$, and $H_{el} = \sum_i \epsilon_i
c_{i}^\dag c_{i}$. Here, $a$, $b_{0}$, and $c_{i} $ are annihilation
operators for exciton, phonon, and electron at the $i$th level,
respectively, while $\epsilon_0$, $\hbar\omega_0$, and $\epsilon_i$ are
their energies. The interactions of exciton with phononic and
electronic excitations can be written as
\[
H_I =
 a^{\dag} a [ V_1  (
b^{\dag}_{0} + b_{0} ) +V_2 b^{\dag}_{0} b^{\dag}_{0}+  V_3 (
b^{\dag}_{0} b^{\dag}_{0}+ b_{0}b_0 )
\]
 \begin{equation}
 +\sum_{i, j}  g_{i,j} c_{i}^\dag
c_{j} ].
 \end{equation}
That is, the emergence of an exciton alters both potentials felt
by electrons and by vibrations, respectively. For electron states the change of potential
leads to violation of orthogonality expressed by matrix elements $g_{i,j}$. For phonons such violation may include
anharmonic effect and is generally described by the linear element $V_1$, the
diagonal and off-diagonal quadratic elements $V_2$ and $V_3$.
It should be pointed out that the coupling
strengths $g_{i,j}$ and $V_{1,2,3}$ can be tuned by the density of
excitons, or by the power of optical pumping.


In the initial state after pumping an exciton, we are working in the
Hilbert subspace with one exciton and operator $a^\dag a$ can be
replaced by 1. Thus, the phonon and electron parts in the
Hamiltonian are decoupled. For the phonon part, we use the
following orthogonal transformation
\begin{eqnarray}
  b_0=u b +v b^\dag +w, \nonumber \\
  \label{tr} b_0^\dag =u b^\dag +v b +w,
  \end{eqnarray}
where $b$ and $b^\dag$ are annihilation and creation operators in the new
representation obeying the boson
commutation rules if
$
  u^2-v^2 =1.
$ In this representation the
Hamiltonian is diagonalized when $u$, $v$ and $w$ satisfy equations
\begin{equation}
  u_l = \sqrt{\frac{\epsilon_p+\epsilon_r }
  { 2 \epsilon_r}}, \,
   v_l = \pm \sqrt{\frac{\epsilon_p-\epsilon_r}
   { 2 \epsilon_r}},\, w_l = -\frac{V_1}{ \epsilon_p +2V_3},
\end{equation}
where $\epsilon_p = \hbar\omega_0 + V_2$, $\epsilon_r =
  \sqrt{\epsilon_p^2 -4 V_3^2}$, and $l=\pm$ represents two
irreducible sets of renormalized phonon states. The phonon Hamiltonian becomes
\begin{equation}
 H'_{ph} = \hbar \omega_l
  b_l^\dag b_l
   +E_l,
  \end{equation}
  where
 \begin{equation}
  \hbar\omega_{\pm} = (\epsilon_p^2 \pm 4 V_3^2 )/\epsilon_r,
 \end{equation}
 and
 \begin{equation}
 E_{\pm} = -\frac{
  V_1^2 (\epsilon_p +3 V_3)}{(\epsilon_p +2V_3)^2}
  -\frac{\epsilon_p}{2} +\frac{\epsilon_p^2 \pm 4 V_3^2}
  {2 \epsilon_r}.
  \end{equation}
The elements of transformation matrix between $\{b_l \}$ and $\{ b_0
\}$ representations denoted by
\begin{equation}
  \eta_{\, l,n; n'} \equiv \langle 0 \left| \frac{( b_l)^n( b^\dag_0)^{n'} }
  {\sqrt{n!n'!}} \right|0\rangle
  \end{equation}
can be calculated by the
iteration relation:
\[
  \eta_{ \, l,n;n'} = \frac{\sqrt{n} u_l }{ \sqrt{n'}} \eta_{ \,
  l,n-1;n'-1} + \frac{\sqrt{n+1} v_l }{ \sqrt{n'}} \eta_{ \,
  l,n+1;n'-1}
\]
\begin{equation}
 \label{ite}
 +\frac{w_l }{ \sqrt{n'}} \eta_{ \, l,n;n'-1},
  \end{equation}
with a restriction condition
  \begin{equation}
  \eta_{ \, l,n;n'}=0 \text{   for   } n' < n, \, \, n <0, \text{
  or  } n'<0,
  \end{equation}
and an initial value
 $
  \eta_{ \, l,0;0}
  $ determined by the normalization condition $ \sum_{n'} |\eta_{ \,
  l,0;n'}|^2 =1$.

The electron Hamiltonian can also be diagonalized with an orthogonal
transformation
 $
  c_i =\sum_{i'} q_{i,i'} d_{i'},
 $
\begin{equation}
    H'_{e} = \sum_i \epsilon'_i d^\dag_i d_i,
    \end{equation}
where $\epsilon'_i$ with $i=1,2,\ldots, $ are the renormalized
electron levels. The elements $q_{i,i'}$ and energies $\epsilon'_i$
depend on $\{ \epsilon_i \}$ and on $\{g_{i,j}\}$.
Consider a general electron level structure including hydrogen-like donor levels
and the empty conduction band described by
\begin{equation}
\label{ppp}
  \epsilon_i = \left\{ \begin{array}{l}  -\frac{\epsilon_b}{i^2}, \text{
   for  } i \leq M, \\
   \delta_e (i-M-1),  \text{  for  }  i \geq M+1,
   \end{array} \right.
   \end{equation}
where $\epsilon_b$ is the binding energy of the hydrogen-like donor,
$\delta_e$ is the energy spacing in the conduction band which is
infinitesimal but set to be finite in numerical calculations, and
$M$ is the number of discrete levels. Considering that the coupling
strength is proportional to the overlapping of the wave functions,
we adopt the following expression for the matrix elements
\begin{equation}
  g_{i,j} = \left\{ \begin{array}{l} g_1 (ij)^{-3/2}, \text{  for  }
  i ,j \leq M, \\ g_2 i^{-3/2}, \text{  for  } i \leq M \text{  and
  } j > M,
  \end{array} \right.
  \end{equation}
where $g_1$ and $g_2$ are adjustable parameters describing the
coupling strengths of a hydrogen-like state with another one
and with a state in conduction band,
respectively.

Finally the exciton-electron-phonon many-body wave functions can be
written as
\begin{equation}
\label{ent}
  \Psi_{i,l,n} =
   \frac{a^{\dag} (b_l^{\dag} )^{n}
   d^\dag_{i} }{\sqrt{n!}} | 0 \rangle ,
\end{equation}
where $|0\rangle $ denotes the vacuum and $n$ is the number of phonons. This state
has eigen energy
\begin{equation}
 E_{i,l,n} = \epsilon_0 +E_l +n\hbar
 \omega_l + \epsilon'_i.
 \end{equation}

The statistical distribution $F_{i,l,n}(k_BT)$ of $\Psi_{i,l,n}$ depends on the non-equilibrium pumping and
relaxation processes. Under the condition of local equilibrium, the distribution function is
\begin{equation}
 F_{i,l,n }(T) = \frac{1}{Z} \exp \left( -\frac{E_{i,l,n}}{k_BT} \right) ,
 \end{equation}
where $Z$ is the partition function
\[
 Z= \sum_{i,l, n} \exp \left( -\frac{E_{i,l,n}}{k_BT} \right).
 \]

Light-exciton coupling described by $H_R = \sum_k W (f_k^\dag a
+\text{H.c.})$ with $f^\dag_k$ being the creation operator of photon
$k$ leads to annihilation of exciton part in state $\Psi_{i,l,n}$
and to creation of a photon. If the transition begins at $t=0$, the
time dependent initial wave function can be written as
\begin{equation}
  \psi^{int}_{i,l,n}(t) = \Psi_{i,l,n} e^{-\text{i} E_{i,l, n} t -\Gamma
 t},
 \end{equation}
where $\Gamma$ describes the decaying process of state
$\Psi_{i,l,n}$ due to the annihilation of the exciton. The final state with a photon at mode $k$ is
\begin{equation}
  \psi^{fin}_{k,i',n'}(t) =\frac{f^{\dag}_k (b_0^{\dag} )^{n'}
   c^\dag_{i'} }{\sqrt{n'!}}e^{-\text{i} (
 \hbar \omega_k +n'\hbar \omega_0 + \epsilon_{i'} ) t } | 0 \rangle ,
 \end{equation}
where $\omega_k$ is the photon frequency. The optical matrix
element between $\psi^{int}_{i,l,n}(t)$ and
$\psi^{fin}_{k,i',n'}(t)$ is
\begin{equation}
 \alpha_{i,l,n;k,i',n'} \propto \overline{ \langle \psi^{fin}
 _{i,l,n}(t)| H_l | \psi^{ini}
 _{k,i',n'}(t) \rangle },
 \end{equation}
where the line over the term on the right hand side denotes the
averaging over time. Note that the relative phases of these
components are coherent and can be inherited by the emitted photons,
the resultant spectra can exhibit interfering structures. So we have
\begin{equation}
\label{alpha}
 \alpha_{i,l,n;k,i',n'}\propto \frac{ \eta_{\, l,n;n'} q_{i,i'}}{\text{i}[ \hbar
 \omega_k -(E_{i,l,n} -n' \hbar \omega_0 -\epsilon_{i'})] -\Gamma}
 ,
\end{equation}
and the spontaneous emission intensity can be written as
 \begin{equation}
 \label{coh}
 \rho (\omega, T) = \sum_{i,l,n} \sum_k F_{i,l,n}(T)
  \left|
 \sum_{i',n'} \alpha_{i,l,n; k,i',n'}  \right|^2 \delta(\omega_k
 -\omega).
 \end{equation}

In Eq. (\ref{coh}) effects of phononic and electronic excitations
and possible coherence between them are reflected as illustrated in
the left panel of Fig. 1. As an application we calculate the
emission spectrum of ZnO and the result is shown in the right panel
of Fig. 1. By comparison with the measured spectrum for ZnO
\cite{exp}, a global agreement between the theory and the experiment
is achieved. The main peak corresponding to the contribution from
the components with zero phonon and with the lowest electron state $(n'=0, i'=1)$ in a group of low-lying
many-body states $ \{\Psi_{i\in {\cal P},l=2,n=0}\}$ is denoted as
A. Its first-order and second-order phonon replicas are C and E,
contributed from the components with indices $(n'=1,i'=1)$ and
$(n'=2,i'=1)$, respectively, with energy separations exactly
equal to $\hbar \omega_0$. In this group of many-body states, the
energies of the electron part is close to the lowest one, i.e.,
$\epsilon'_{i\in {\cal P}} -\epsilon'_1 < k_BT$, so the statistical
weights from them are not negligible. Another set of the zero-phonon
and one-phonon Stokes lines, also with exact energy separation
$\hbar \omega_0$, are formed from the components with indices $i' >
1$ and $n'=0,1$ in the same group of many-body states, and indicated
as B and D in Fig. 1. This set is associated with electron
excitations from the lowest states to the excited states including the states in the conduction band. 
From Eqs. (\ref{ppp},\ref{alpha},\ref{coh}) we can
see that set B-D is below set A-C by an energy of the order of
$\epsilon_b$ of the hydrogen-like states. On the high-energy side of
the zero-phonon line (ZPL) there appear some weak anti-Stokes lines G
and H originating from many-body states with higher energies. In
previous theories
 they are usually attributed to the radiation accompanied by absorption of phonons.

\begin{figure}[ht]
\includegraphics[width=8.6cm]{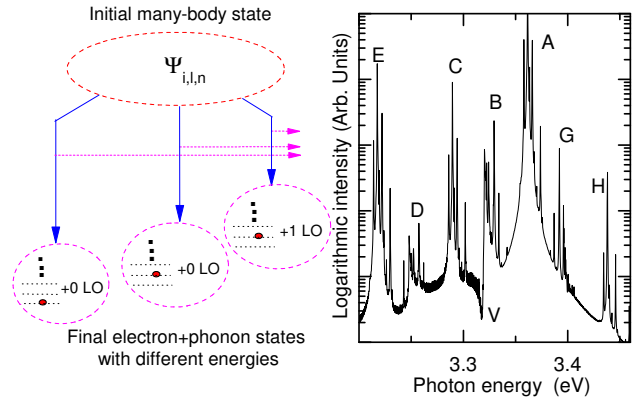}
\caption {(Color online) Left panel: Illustration of transitions from
a many-body state with an exciton to various final
electron and phonon states without exciton. Right panel:
Calculated spontaneous emission spectrum of ZnO taking into account both the phonon and
electron excitations. The parameters are: $\epsilon_0 =3.3676$eV,
$\hbar\omega_0 =0.072$eV, $V_1=0.012$eV, $V_2 = 0.014 $eV,
$V_3=0.02$eV, $g_1=-0.022$eV, $g_2 = 0.00036$eV, $\epsilon_b
=0.038$eV, $ \delta_e =0.0005$eV, $M=6$, $\Gamma = 0.0001$eV, and the
number of included conduction band states is 300. The equivalent
temperature for the local equilibrium distribution is $k_BT =
27.4$meV, which may be higher than the environment temperature due to
the pumping. } \label{fig1}
\end{figure}

In the following we
discuss several important characteristics in the spectrum.
If the coupling to electrons and the anharmonicity of phonons are set to zero, the theory reduces
to the well-known Huang-Rhys theory \cite{huang} as shown in Fig. 2(a), giving an important support to the
present treatment of interactions. When the exciton-electron couplings are not zero, new structures appear in
the spectrum. In Fig. 2(b) we plot the spectra of two cases: (i) $g_1\neq 0$ and $g_2=0$ (dotted line); (ii) $g_1\neq 0$ and $g_2\neq 0$ (dashed line). In case (i), the exciton is only bound to localized electron excitations associated with impurity levels. The most pronounced spectral feature is the appearance of the so-called two-electron satellites on the lower side of the ZPL as seen from the peaks of the dotted curve in energy range of $3.32-3.34$eV. The positions of these satellite peaks depend on both the donor energy levels and the coupling strength with the exciton. Another effect of $g_1$ is to cause the splitting of phonon lines including the ZPL in the Huang-Rhys spectrum. This splitting is often observed in fine experimental spectra as shown later. In case (ii) a coupling to the continuous excitations associated with the conduction band is further included. Such a coupling produces a continuum background in the spectrum as can be seen from the dashed line in Fig. 2(b). It is more interesting that a distinct Fano dip is created between the one-phonon line and the above-mentioned two-electron satellites. In Fig. 2(c) we show the effect of the anharmonicity in the exciton-phonon coupling. One most direct effect of the anharmonic coupling is the enhancement of the 2-phonon line. Another effect is the new structures appearing on the higher energy side of the ZPL, causing the breaking of the mirror symmetry of the phonon Stokes and anti-Stokes lines about the ZPL. Note that the anharmonic coupling also gives rise to a shift of the whole spectrum due to the change of eigenenergies of many-body states. In Fig. 2(d) the temperature dependence of the Fano dip is depicted. It is clear that increasing temperature can destroy the quantum coherence, resulting in the suppression of the Fano lineshape. This trend has already been observed in experiment \cite{exp}.

\begin{figure}[ht]
\includegraphics[width=8.6cm]{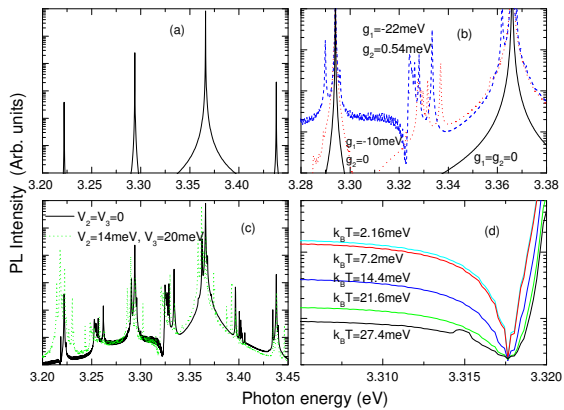}
\caption {(Color online) Calculated emission spectra for different coupling cases. Remaining parameters are the same as those in Fig. 1. (a) $V_2=V_3=g_1=g_2=0$; (b) Different exciton-electron couplings in the case of $V_2=V_3=0$; (c) Different anharmonicity parameters $V_2$ and $V_3$; (d) Evolution of the Fano dip by increasing the temperature.  } \label{fig2}
\end{figure}

In Fig. 3 we present a direct comparison between fine experimental spectra and theoretical curves in two energy regions of the ZPL (left panel) and two-electron satellites (right panel). In both the regions a general agreement between theory and experiment is obtained. Especially, most fine structures in the experimental curves can be reproduced in the theoretical results. This leads to a new understanding of the fine structures in the experimental spectra.

\begin{figure}[ht]
\includegraphics[width=8.6cm]{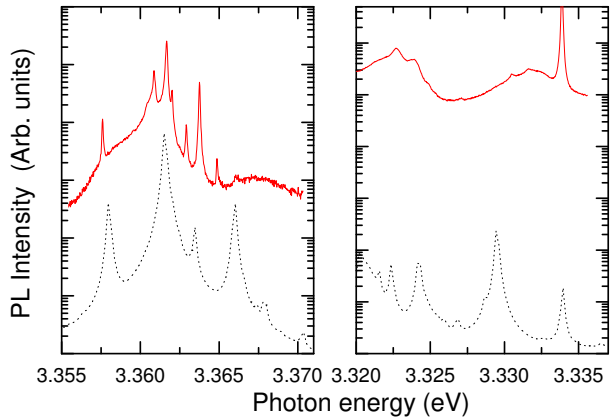}
\caption {(Color online) Comparison between fine experimental ZnO spectra (red solid lines) and theoretical curves (black dotted lines) in two energy regions of the ZPL (left panel) and two-electron satellites (right panel).  } \label{fig3}
\end{figure}

In summary, a microscopic theory for the spontaneous emission of excitons simultaneously coupled to both phonon and electron excitations is developed on the basis of direct solutions of the exciton-electron-phonon composite states. The calculated spectrum using the theory is
quantitatively consistent with the experimental spectrum of ZnO. Most fine structures in the experimental spectrum have been identified.

{\it Acknowledgments} This work was supported in Hong Kong by Hong Kong RGC-CERG
Grant (No. HKU 7056/06P), in Nanjing by National Foundation of
Natural Science in China Grant Nos. 60276005 and 10474033, and by
the China State Key Projects of Basic Research (2005CB623605). The authors wish to thank L. Ding, J. Q. Ning, X. M. Dai, and J. N. Wang for their contributions to the measurements of fine photoluminescence spectra of ZnO. We also thank C. C. Ling for providing ZnO samples.


\end{document}